\documentclass[doublespacing]{elsart}

\usepackage{graphics}
\usepackage{graphicx}
\usepackage{epsfig}
\usepackage{amssymb}

\begin{document}

\begin{frontmatter}



\title{Prisoner's Dilemma on community networks }


\author[lable1,label2]{Xiaojie Chen},
\author[lable1,label2]{Feng Fu},
\author[lable1,label2]{Long Wang\corauthref{cor}}
\corauth[cor]{Corresponding author} \ead{longwang@mech.pku.edu.cn}
\address[lable1]{Intelligent Control Laboratory, Center for Systems and Control, Department of Mechanics and Space Technologies, College of Engineering, Peking University, Beijing 100871, China}
\address[label2]{Department of Industrial Engineering and Management, College of Engineering, Peking University, Beijing 100871, China}
\begin{abstract}
We introduce a community network model which exhibits scale-free
property and study the evolutionary Prisoner's Dilemma game (PDG)
on this network model. It is found that the frequency of
cooperators decreases with the increment of the average degree
$\bar{k}$ from the simulation results. And reducing
inter-community links can promote cooperation when we keep the
total links (including inner-community and inter-community links)
unchanged. It is also shown that the heterogeneity of networks
does not always enhance cooperation and the pattern of links among
all the vertices under a given degree-distribution plays a crucial
role in the dominance of cooperation in the network model.
\end{abstract}

\begin{keyword}
Community networks, Prisoner's Dilemma, Cooperation, Heterogeneity
\PACS 89.75.Hc, 02.50.Le, 87.23.Ge
\end{keyword}
\end{frontmatter}

\section{Introduction}

Cooperation is an essential ingredient of evolution. Understanding
the emergence and persistence of cooperation among selfish players
in evolution is one of the fundamental and central problems. As
one typical game, the Prisoner's Dilemma game (PDG), has become a
world-wide known paradigm for studying the emergence of
cooperative behavior between unrelated individuals. In the
original PDG, two players simultaneously decide whether to
cooperate or defect. The defector will always have the highest
reward $T$ (temptation to defect) when playing against the
cooperator which will receive the lowest payoff $S$ (sucker
value). If both cooperate they will receive a payoff $R$ (reward
for cooperation), and if both defect they will receive a payoff
$P$ (punishment). Moreover, these four payoffs satisfy the
following inequalities: $T >R >P >S$ and $T+S<2R$. It is not
difficult to recognize that it is best to defect for rational
players to get the highest payoff independently in a single round
of the PDG, but mutual cooperation results in a higher income for
both of them. Therefore, this situation creates the so-called
dilemma for selfish players.

To find under what conditions the cooperation emerges on the PDG,
various mechanisms of enforcing cooperation have been explored
\cite{1}. Departure from the well-mixed population scenario, Nowak
and May introduced a spatial evolutionary PDG model in which
players located on a lattice play with their neighbors \cite{2}.
In each round, players adopt the strategy of their most successful
neighbors' in term of their payoff. It has been shown that the
spatial effect promotes substantially the emergence of
cooperation.

In the past few years, the evolutionary PDG has been studied on
different network models such as small-world structure \cite{3},
regular and random graphs by using other mechanisms to enhance
cooperation \cite{4,5,6,7,8}. In all these models, each player
occupies one vertex of the networks. The edges denote links
between players in terms of game dynamical interaction. Each
player just interacts with its adjacent players. Santos \textit{et
al}. have studied the PDG and snowdrift game (SG) on scale-free
networks \cite{9} and found that cooperation dominates in both the
PDG and the SG, for all values of the relevant parameters of both
games \cite{10}. Their results show that the heterogeneity of
networks favors the emergence of cooperation. In addition, Much
attention has been given to the interplay between evolutionary
cooperative behavior and the underlying structure \cite{11,12}.

In this paper, we focus on the PDG on community networks which can
reflect a lot of real-world complex networks such as social and
biological networks. The community network model and update rule
for the PDG are introduced in Sec. 2. Simulation results are shown
for some parameters and their corresponding explanations are
provided in Sec. 3. At last, conclusions are made in Sec. 4.

\label{}


\section{The Model}

We first construct the evolving network model which exhibits
community structure and scale-free properties \cite{13}. We assume
that there is a total of $M$ ($M\geq 2$) communities in the
network model. The evolving model is defined by the following
steps: (1) Starting from $m_0$ ($m_0>1$) fully connected vertices
in each community; and at the same time, there is a
inter-community link between every two different communities. The
vertex to which the inter-community links connect are selected
fixedly in each community (the red points in Fig. 1). (2) A new
vertex is added to a randomly selected community at each time
step. The new vertex will be connected to $m$ ($1\leq m\leq m_0$)
existing vertices in the same community through $m$
inner-community links, and to $n$ ($0\leq n\leq m$) existing
vertices in other $M-1$ communities through inter-community links
(growth). (3) When choosing the vertices to which the new vertex
connects in the same communities through inner-community links,
one assumes that the probability $p_{ij}$ that a new vertex will
be connected to node $i$ in community $j$ depends on the degree
$l_{ij}$  of that vertex: $p_{ij}=l_{ij}/\sum_{k} l_{kj}$ ; When
choosing the vertices to which the new vertex connects in the
other communities through inter-community links, one assumes that
the probability $p_{ik}$ that a new vertex will be connected to
node $i$ in the community $k$ ($ k \neq j$) also depends on the
degree $l_{ik}$ of that vertex: $p_{ik}=l_{ik}/\sum_{i, k, k\neq
j} l_{ik}$ (preferential attachment). After $t$ time steps this
algorithm produces a grape with $N=Mm_0+t$ vertices and
$[Mm_0(m_0-1)+M(M-1)]/2+(m+n)t$ edges, in which older vertices in
each community in the generation process are those which tend to
exhibit larger values of the connectivity. And the average degree
$\bar{k}$ of the networks model is about $2(m+n)$. For example,
Fig. 1 shows an initial network with $M=3$ and $m_0=3$. The
degree-distribution of the network model is shown in Fig. 2, and
we observe that the degree distribution obeys a power-law form.
\begin{figure}
\centering
\includegraphics[scale=0.9]{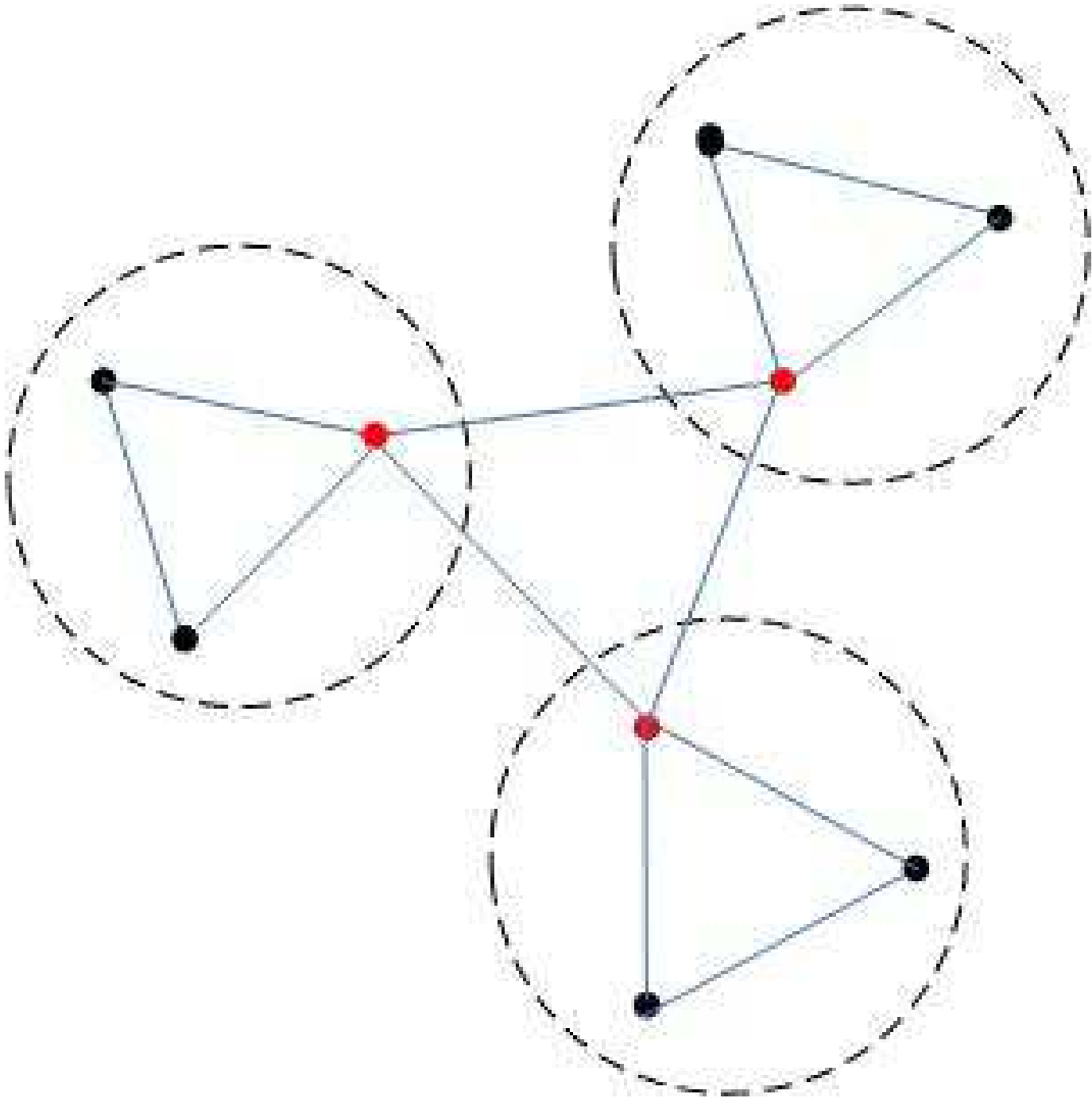}
\caption{The sketch graph of the model with $M=3$ communities and
$m_0=3$. The red dots are chosen to connect to each other between
every two different communities.}
\end{figure}
\begin{figure}
\centering
\includegraphics[scale=0.5]{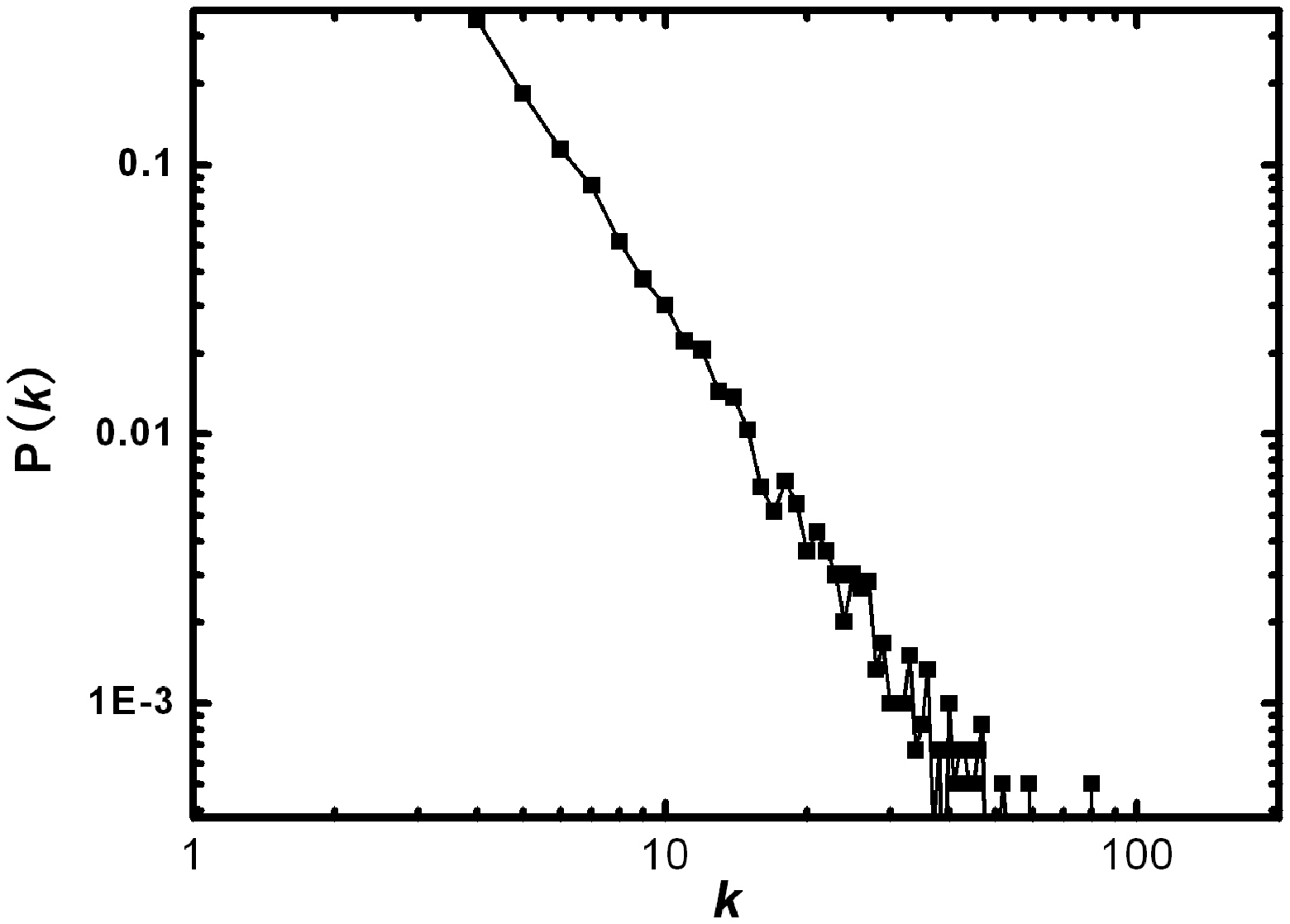}
\caption{The degree distribution of a network with $N=6000$,
$M=3$, $m_0=3$, $m=3$ and $n=1$. }
\end{figure}

The model is generated via growth and preferential attachment,
then we set up a system of $N$ players arranged at the vertices of
this network model. Each player who is a pure strategist can only
follow two simple strategies: C (cooperate) and D (defect). In one
generation, each player plays a PDG with its neighbors. Let's
represent the players' strategies with two-component vector,
taking the value $s=(1,0)$ for C-strategist and $s=(0,1)$ for
D-strategist. The total payoff of a certain player $x$ is the sum
over all interactions, so the payoff $P_x$ can be written as
\begin{equation}
P_x=\sum_{y\in \Omega_x} s_x As_y^T, \label{eq1}
\end{equation}
where $\Omega_{x}$ is the set of neighbors of element $x$ and $A$
is the payoff matrix.
\begin{equation}
A=\left[\matrix{R& S \cr T &P}\right]. \label{eq2}
\end{equation}
Introduced by Nowak and May \cite{2,14}, the payoff matrix can be
written as a simplified version
\begin{equation}
A=\left[\matrix{1 &0 \cr b&0}\right],\label{eq3}
\end{equation}
where $b$ represents the advantage of defectors over cooperators
and $1<b<2$. Therefore, we can rescale the game depending on the
single parameter $b$.

After this, the player $x$ will inspect the payoff collected by
its neighbors in the generation, and then update its strategy for
the next generation to play by the following rule \cite{10}. It
will select one player $y$ randomly from its neighbors. Whenever
$P_y>P_x$, player $x$ will adopt the strategy of player $y$  with
probability given by:
\begin{equation}
W_{s_x\leftarrow s_y}=(P_y -P_x)/(Dk_>),\label{eq4}
\end{equation}
where $k_>=max\{k_x, k_y\}$ and $D=T-S$. And $k_x$ and $k_y$ are
the connectivity numbers of player $x$ and $y$, respectively. We
use a synchronous update, where all the players decide their
strategies at the same time. All pairs of players $x$ and $y$ who
are directly connected on the network model engage in each
generation of the PDG by using the update rule of Eq.~(\ref{eq4}).

\label{}
\section{Simulations and Discussion}
In the following, we show the results of simulations carried out
for a population of $N=6000$ players occupying the vertices of the
networks with $M=3$ same size of communities. The initial
strategies (cooperators or defectors) are randomly distributed
with equal probability. Then more than $11000$ generations are
played to allow  for equilibrium frequency of cooperators and
defectors which are achieved by averaging over the last 1000
generations. The simulation results of the frequency of
cooperators as a function of $b$ for the PDG has been shown.
Moreover, each data point averages over 40 realizations of both
the networks and the initial conditions.
\begin{figure}
\centering
\includegraphics[scale=0.6]{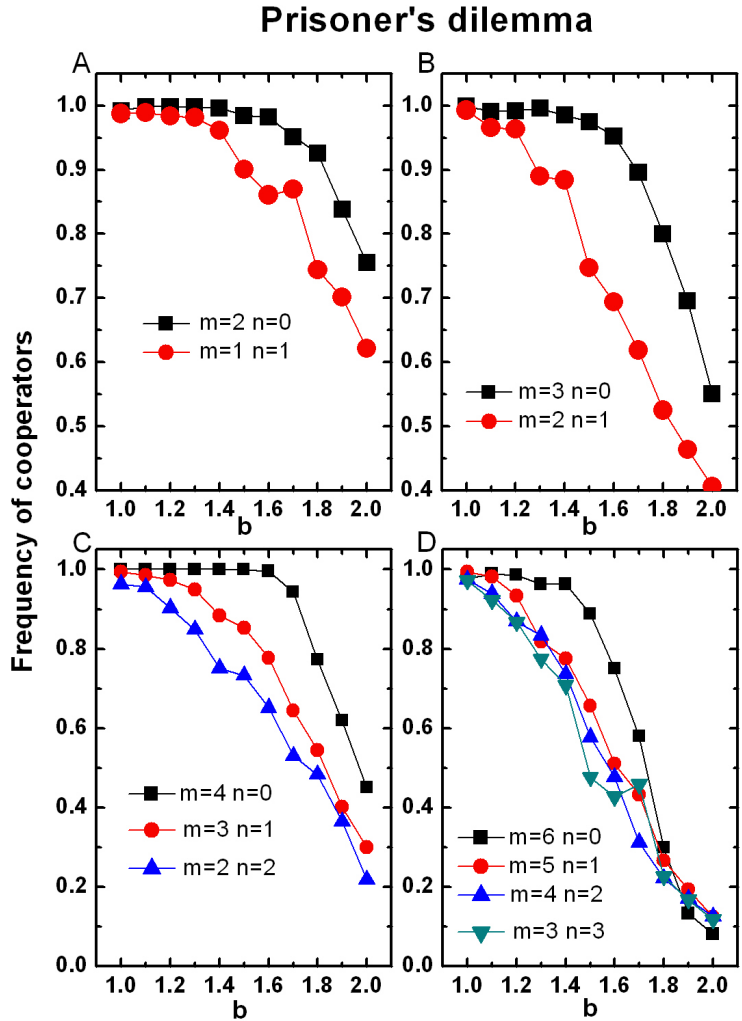}
\caption{Frequency of cooperators for the PDG as a function of the
parameter $b$ for different values of the average degree
$\bar{k}$. A, $m_0=2$ and $\bar{k}=4$; B, $m_0=3$ and $\bar{k}=6$;
C, $m_0=4$ and $\bar{k}=8$; D, $m_0=6$ and $\bar{k}=12$. The
colored lines in each subgraph  correspond to different $m$ and
$n$ for a fixed value of $\bar{k}$, respectively.}
\end{figure}
\begin{figure}
\centering
\includegraphics[scale=0.6]{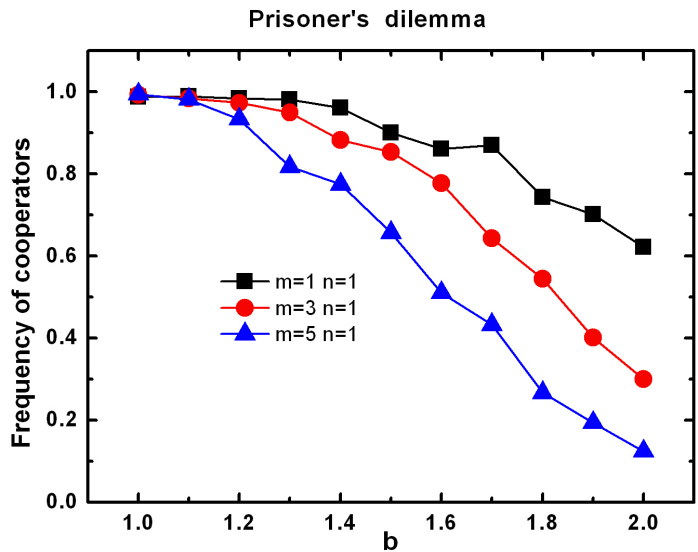}
\caption{Frequency of cooperators in the PDG as a function of the
parameter $b$ for different values of $m$, given a fixed value of
$n=1$.}
\end{figure}
\begin{figure}
\centering
\includegraphics[scale=0.6]{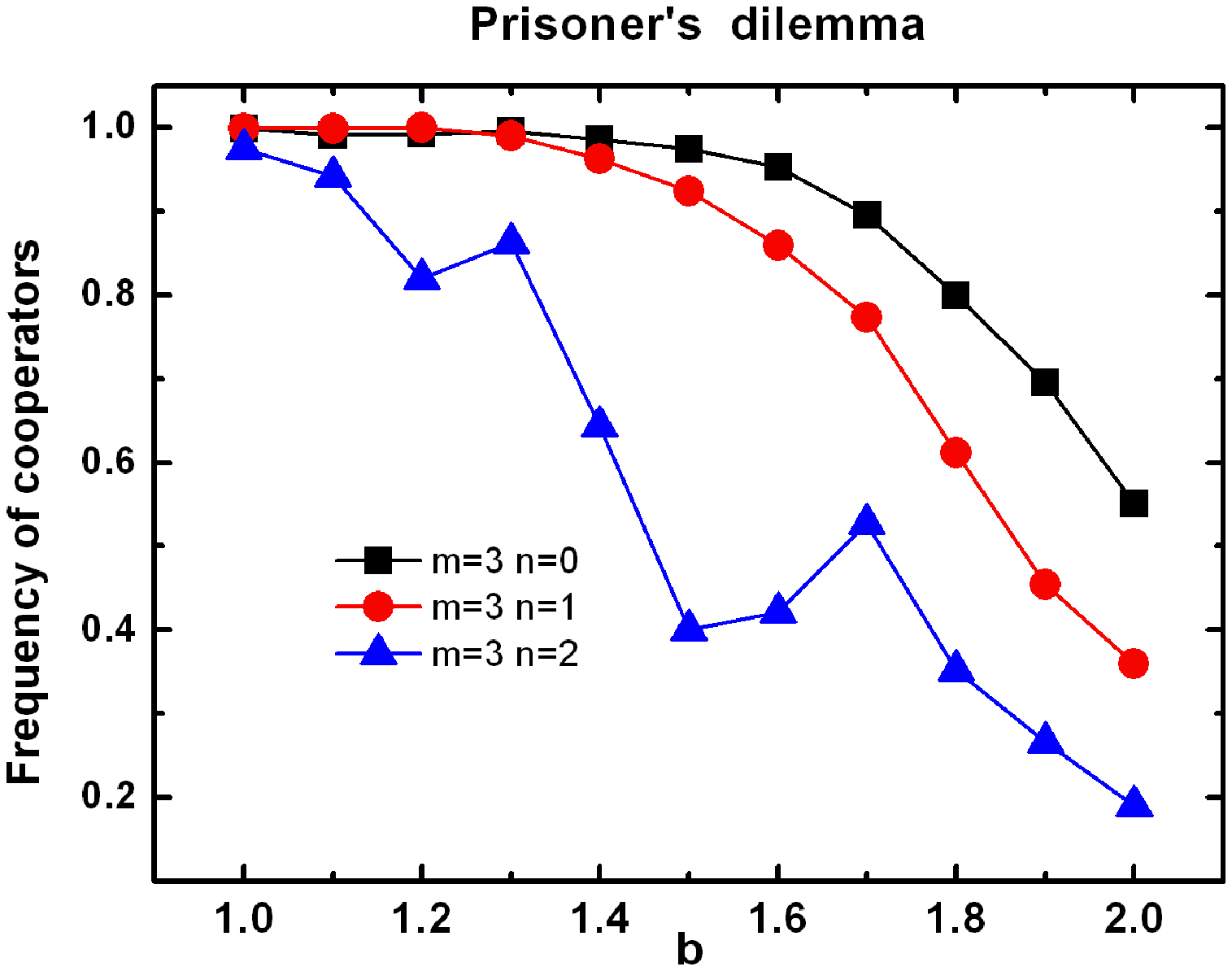}
\caption{Frequency of cooperators in the PDG as a function of the
parameter $b$ for different values of $n$, give fixed values of
$m=3$ and $m_0=3$.}
\end{figure}

In Fig. 3, we show the results for the PDG on the network model
for different values of the average degree $\bar{k}$. It shows
that the higher the value of the average degree $\bar{k}=2(m+n)$,
the more unfavorable cooperation becomes, especially for a large
value of the parameter $b$. From Fig. 3A to Fig. 3D, we have found
that the larger the value of the parameter $m/n$ is, the more
favorable cooperation becomes over the large range of $b$ when the
average degree $\bar{k}$ is unchanged. Moreover, for small $b$,
cooperation dominates in the PDG when the value of the parameter
$\bar{k}$ is small; however, the frequency of cooperators
decreases rapidly over the entire range of the parameter $b$ when
the value of the parameter $\bar{k}$ is high. Fig. 4 shows the
frequency of cooperators as a function of $b$ with different $m$
when we keep $n=1$. The frequency of cooperators decreases when
the value of the parameter $m$ increases for a given fixed $b$.
Fig. 5 shows the frequency of cooperators as a function of $b$
with different $n$ when we keep $m=3$. The frequency of
cooperation decreases when the value of the parameter $n$
increases for a given fixed $b$.

From the simualtion results, we have found that the smaller the
value of the parameter $\bar{k}$, the more favorable cooperation
becomes; besides, the smaller the value of the parameter $m/n$,
the more unfavorable cooperation becomes for a given fixed
$\bar{k}$. These results can be explained in the following ways.
The average degree $\bar{k}$ of the network models can affect the
frequency of cooperators \cite{10,15}. On this community network
model, the small value of  $\bar{k}$ is benefit to cooperation. On
the one hand, the heterogeneity of the network structure can
promote cooperation; the direct inter-connections of hubs play an
significant role in enhancing cooperation \cite{10,16,17,18}. On
this network model, the initial-fixed vertices connected to
between each community are expected to be the largest hubs in each
community. We have found that these fixed vertices are always hubs
with largest degree in each community for small $\bar{k}$ by data
analysis. However, if we increase the value of $m$ or $n$, the
expected hubs may not have the largest connectivity any more,
though the degree distribution still obeys a power-law form.
More and more older vertices would compete to be the hubs for high
$\bar{k}$, and then the number of direct links among hubs would
decreases. It is because if there are no direct links connecting
to any two vertices belonging to different communities initially,
they would become unconnected forever. Therefore, large $m$ or $n$
leads to the case that the number of direct links among hubs
decreases. This would inhibit cooperation since the number of
direct links among larger hubs from different communities
decreases. On the other hand, higher connectivity should reduce
cooperation; in particular, more cooperation should emerge if
connectivity is low \cite{15}. Our simulation results confirm this
conclusion. When we increase the value of $m$ or $n$, players will
have more neighbors, but having more neighbors for players does
not pay at all. Beacuse it would lead the community network to the
high connectivity, and then the cooperative behaviour becomes
inhibited.

It is well-known that scale-free networks promote cooperation
\cite{10}. And scale-free networks have most of their connectivity
clustered in a few vertices, like the amplifier structures, such
as loops and circulations which are potent for cooperation
\cite{19}. While for a given fixed $\bar{k}$, the best one for
promoting cooperation is to keep $n=0$. In this situation, it
apparently becomes to be similar to three hubs which are connected
to each other. And all the newly growing vertices only connect to
other vertices belonging to the same community. Each community is
a standard BA scale-free network, so the whole network consists of
amplifier structures to promote cooperation. As the ratio m/n
decreases, the inner-community links number decreases and the
inter-community links number increases. Due to the community
structure, in general, the vertices in different communities may
not have direct links connecting to each other, then these
vertices from different communities do not form loops through
inter-community links. As a result, the number of loops on the
community network decreases and the whole network does not favor
cooperation. Therefore, when the ratio $m/n$ decreases for a fixed
$\bar{k}$, the cooperation becomes unfavorable.

Therefore, it is not surprising that the small number of links on
this community networks can promote cooperation by considering
these factors. Community sturcture is an ubiquitous phenomena in
all kinds of complex networks, especially in human society. From
our results, taking account into the effect of community
structure, some certain community structures promote cooperation
greatly; besides, different patterns of links among all the
vertices under a given degree-distribution can affect cooperation,
especially connections between hubs can enhance cooperation
greatly. Moreover, we have confirmed that all the simulation
results are valid for different population size $N$ and community
size $M$.

\section{Conclusions}
To sum up, we have studied the cooperative behavior of the
evolutionary PDG on the community networks and found that reducing
inner-community and inter-community links can promote cooperation.
The heterogeneity of networks is not always positive to enhance
cooperative behavior and the situation of connections among all
the vertices, especially some certain structures and direct
connections between hubs plays a more significant role in the
dominance of cooperation. Graph topology plays a determinant role
in the evolution of cooperation \cite{20}. However, it is
necessary to explore more direct and essential factors that
facilitate cooperation to dominate for future work.

\section*{Acknowledgements}
This work was supported by National Natural Science Foundation of
China (NSFC) under grant No. 60674050 and No. 60528007, National
973 Program (Grant No. 2002CB312200), and 11-5 project (Grant No.
A2120061303).

\newpage

\end{document}